%% file: paper.tex
\documentclass[lettersize,journal]{IEEEtran}
\usepackage{amsmath,amsfonts}
\usepackage{algorithm}
\usepackage{algpseudocode}  
\usepackage{array}
\usepackage[caption=false,font=footnotesize,labelfont=rm,textfont=rm]{subfig}
\usepackage{textcomp}
\usepackage{stfloats}
\usepackage{url}
\usepackage{verbatim}
\usepackage{graphicx}
\usepackage{setspace}
\usepackage{cite}
\usepackage{enumitem}
\usepackage{hyperref}
\usepackage[usenames]{color}

\include{header}

\begin{document}

\title{Efficient Wireless Federated Learning via Low-Rank Gradient Factorization}

\author{Mingzhao Guo, Dongzhu Liu, Osvaldo Simeone, and Dingzhu Wen
\thanks{The work of O. Simeone was partially supported by the European Union’s Horizon Europe project CENTRIC (101096379), by the Open Fellowships of the EPSRC (EP/W024101/1), by the EPSRC project (EP/X011852/1).  The work of Mingzhao Guo and Dingzhu Wen was supported in part by the National Natural Science Foundation of China under Grant No. 62401369 and in part by the Shanghai Sailing Program under Grants No. 23YF1427400. This work was supported in part by the Platform of Computer and Communication provided by ShanghaiTech University.
(Corresponding author: Dongzhu Liu.)}
\thanks{M. Guo and D. Wen are with the School of Information Science and Technology, ShanghaiTech University, China (e-mail: \{guomzh1, wendzh\}@shanghaitech.edu.cn).}
\thanks{D. Liu is with the School of Computing Science, University of Glasgow, UK (e-mail: dongzhu.liu@glasgow.ac.uk).}
\thanks{Osvaldo Simeone is with the King’s Communications, Learning \& Information Processing (KCLIP) lab, Centre for Intelligent Information Processing Systems (CIIPS), Department of Engineering, King’s College London, WC2R 2LS London, U.K. (e-mail: osvaldo.simeone@kcl.ac.uk).}
}

\maketitle

\begin{abstract}
This paper presents a novel gradient compression method for federated learning (FL) in wireless systems. The proposed method centers on a low-rank matrix factorization strategy for local gradient compression that is based on one iteration of a distributed Jacobi successive convex approximation (SCA) at each FL round. The low-rank approximation obtained at one round is used as a ``warm start" initialization for Jacobi SCA in the next FL round. A new protocol termed over-the-air low-rank compression (Ota-LC) incorporating this gradient compression method with over-the-air computation and error feedback is shown to have lower computation cost and lower communication overhead, while guaranteeing the same inference performance, as compared with existing benchmarks. As an example, when targeting a test accuracy of $70\%$ on the Cifar-10 dataset, Ota-LC reduces total communication costs by at least $33\%$ compared to benchmark schemes.
\end{abstract}

\begin{IEEEkeywords}
Wireless federated learning, gradient factorization, over-the-air computation, MIMO. 
\end{IEEEkeywords}

\section{Introduction}

\IEEEPARstart{F}{ederated} Learning (FL) is a distributed machine learning paradigm designed to collaboratively train a shared model across multiple devices without the need to access their raw data \cite{dean2012large}. The deployment of FL at the network edge introduces significant communication overhead due to the need to exchange high-dimensional model updates. This communication challenge becomes even more pronounced in the context of wireless networks operating over shared radio resources.

With noiseless, interference-free, but constrained communication, gradient compression has emerged as an efficient approach for reducing communication overhead. Representative gradient compression methods fall into three main categories, namely quantization, sparsification, and low-rank approximation. Gradient quantization supports digital communication, whereby each entry of the gradient is represented by using a limited number of bits \cite{alistarh2017qsgd}, or even only one bit \cite{bernstein2018signsgd1,karimireddy2019error}, and it can utilize a Grassmann manifold codebook \cite{du2020high} or vector quantization \cite{shlezinger2020uveqfed}.  Gradient sparsification decreases the number of non-zero entries in the communicated gradient, with methods including Top-$K$, Rand-$K$, and threshold-based selection \cite{alistarh2018convergence,stich2018sparsified,sahu2021rethinking}. Compressive sensing (CS) recovery algorithms can further increase the sparsing level, supporting both analog and digital transmission \cite{zhong2023over, becirovic2022optimal, jeon2020compressive}. Low-rank approximation, the focus of this work, uses low-rank factors to approximate the high dimensional gradient. These factors can be derived by projecting the original gradient into low-dimensional subspaces shared by multiple devices, resulting in random linear coding (RLC) \cite{abdi2020analog}, or by employing power iteration to factorize the gradient matrix, as demonstrated by powerSGD \cite{vogels2019powersgd}.

Introducing wireless channels in the implementation of FL presents both novel design challenges and opportunities, particularly in the exploration of over-the-air aggregation in FL updates and in the utilization of the degrees of freedom inherent in multiple input multiple output (MIMO) communications. For a simplified wireless communication model with additive white Gaussian noise channels, \cite{makkuva2023laser} applies powerSGD \cite{vogels2019powersgd} for FL updates, integrating over-the-air computation to enhance communication efficiency. Considering a fading channel model, the authors of \cite{xing2021federated} leverage RLC as a gradient compression method in the context of decentralized FL. To further optimize communication efficiency, reference \cite{zhong2023over} applied sparsification with a CS algorithm, namely turbo-CS \cite{6883198}, for gradient compression. 
Despite successfully reducing communication overhead in wireless FL systems, these works either impose high computational complexity on local devices as  RLC or CS algorithms involving high-dimensional matrix operations or have to compromise learning performance, thus motivating our work.

This work proposes a gradient compression method induced by solving a low-rank matrix factorization problem at each round of FL. The method centers on  Jacobi successive convex approximation (SCA) \cite{6675875}. In each FL round, it utilizes the approximated low-rank gradient factors from the previous round as the ``warm start"  initialization for the next round and reduces to a single Jacobi SCA iteration. Applying this method to fading  MIMO channels together with over-the-air computation and error feedback, the algorithm, termed over-the-air low-rank compression (Ota-LC), is shown via experiments to outperform existing methods in terms of communication overhead, while also reducing the computational complexity.

\section{System Model}
\label{sec: system model}

\begin{figure}[!htbp]
    \centering
    \includegraphics[width = 1 \linewidth,trim = 0.3cm 0.3cm 0.3cm 0.2cm, clip]{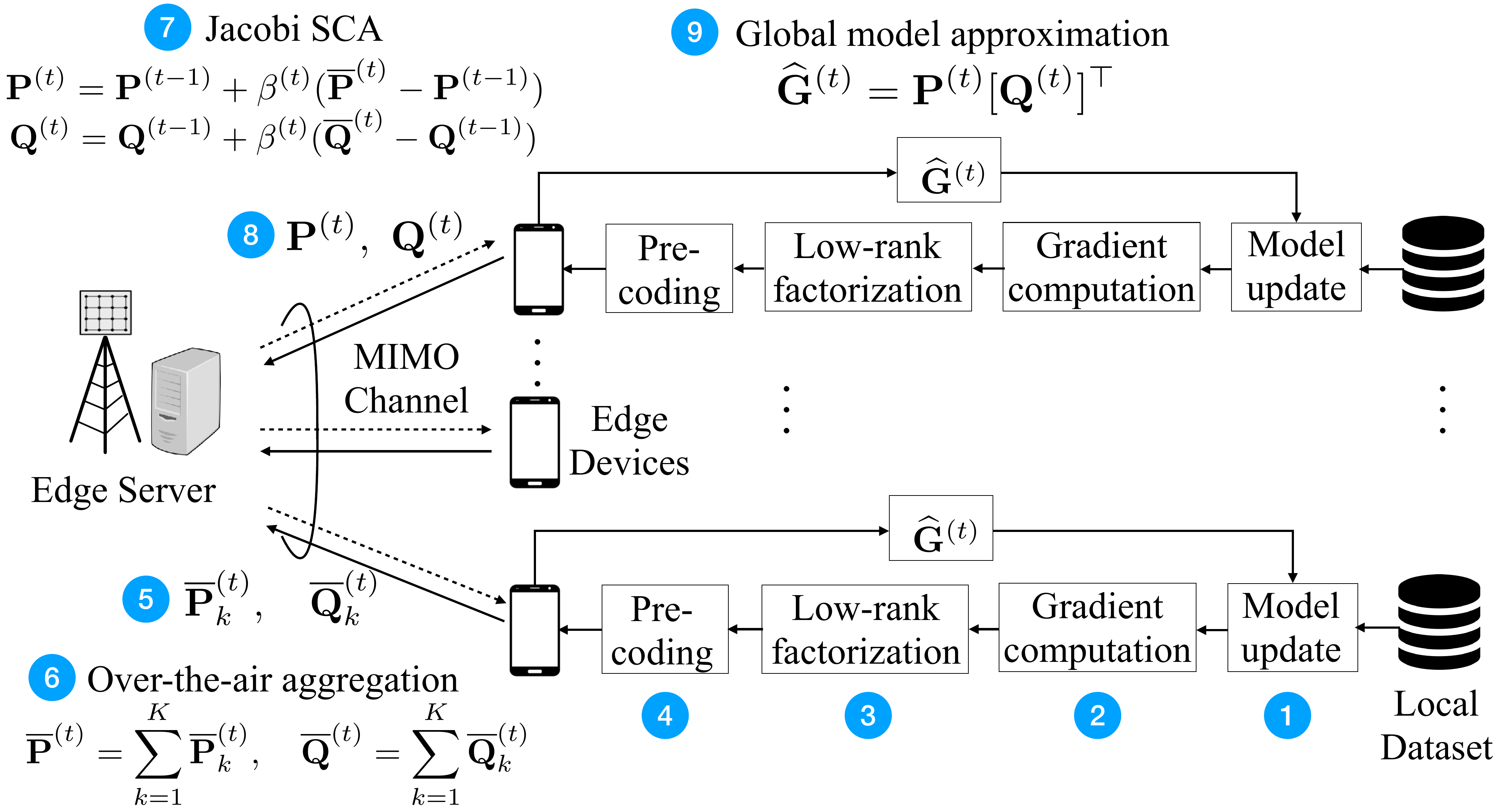}
    \setlength{\abovecaptionskip}{0.cm}
    \caption{Low-rank gradient compression for a MIMO wireless federated learning system.}
    \label{fig:model}
\end{figure}

As shown in Fig. \ref{fig:model}, we consider a federated edge learning (FEEL) system comprising a single edge server and $K$ edge devices connected via a shared MIMO channel.  Each device $k$ has its local dataset $\mathcal{D}_k$.  This consists of labeled data samples $\{(\bu,v)\} \in \mathcal{D}_k$, where $\bu$ denotes the covariates and $v$ denotes the associated label, which may be continuous or discrete.  A tensor $\bW$ is collaboratively trained by the edge devices through communications via the edge server.

\subsubsection{Federated Gradient Descent}\label{sec: FL protocol}
The local loss function for the $k$-th device evaluated at parameter tensor  $\bW$ is given by

\begin{equation}
\label{eq: local loss}
    F_k(\bW)=\frac{1}{{D}_k}\sum\nolimits_{(\bu,v)\in\mathcal{D}_k}f(\bW;\bu,v),
\end{equation}
where $f(\bW;\bu,v)$ is the sample-wise loss function quantifying the prediction error of the model $\bW$ on the training sample $\bu$ with respect to (w.r.t.) its ground-truth label $v$; and $D_k=|\mathcal{D}_k|$ is the cardinality of data set $\mathcal{D}_k$. Accordingly, the global loss function is given as 
\begin{equation}\label{eq: global loss}
 F(\bW)=\frac{1}{D}\sum\nolimits_{k=1}^K D_k F_k(\bW),
\end{equation}
where $D=\sum_{k=1}^KD_k$. The federated learning process aims to minimize the global loss function as
\begin{equation}\label{eq: LP}
\bW^*=\arg\min F(\bW).
\end{equation}


At each $t$-th communication round, by using the current model $\bW^{(t)}$ and the local dataset $\mathcal{D}_k$, each device computes the gradients of the local loss function in \eqref{eq: local loss}, that is
\begin{equation}\label{eq: local gradient}
\nabla F_k (\bW^{(t)})=\frac{1}{D_k}\sum\nolimits_{(\bu,v)\in\mathcal{D}_k}\nabla f\l(\bW^{(t)};\bu,v\r).
\end{equation}
 The devices transmit information about the local gradient in \eqref{eq: local gradient} over a shared wireless channel to the edge server.  Based on the received signal, the edge server obtains the global gradient 
\begin{equation}\label{eq: global gradient}
\bG^{(t)}=\frac{1}{D}\sum\nolimits_{k=1}^K D_k \nabla F_k (\bW^{(t)}) = \sum\nolimits_{k=1}^{K}  \bG_{k}^{(t)},
\end{equation}
where $\bG_{k}^{(t)} = \frac{D_k}{D} \nabla F_k(\bW^{(t)}) $ is the local gradient for device $k$. The edge server updates the model via gradient descent 
\begin{equation}\label{eq: Model update}
\bW^{(t+1)}=\bW^{(t)}-\eta \bG^{(t)},
\end{equation}
where $\eta$ denotes the learning rate. The updated parameter tensor \eqref{eq: Model update} is sent back to all devices on the downlink. As in much of previous works such as \cite{becirovic2022optimal,zhong2023over,xing2021federated}, we assume that downlink communication is ideal,  which is practically well justified when the edge server communicates with a less stringent power constraint than the devices. The steps in \eqref{eq: local gradient}-\eqref{eq: Model update} are repeated until a convergence condition is met, e.g., until the condition $\|{\bf G}^{(t)}\|\leq \epsilon$ is satisfied for some value of $\epsilon>0$ or until approaching the maximum number of communication rounds.

\subsubsection{Communication Model}
We now introduce the uplink wireless MIMO communication model for the FL system.
Each device is equipped with $N_t$ transmitted antennas, and the server is equipped with $N_r$ received antennas. In the $t$-th communication round,  all devices communicate their local gradients through a block-fading channel simultaneously.  We assume the channel state information (CSI) remains constant within $\tau$ communication blocks. The received signal $ \mathbf{Y}^{(t)}\in\mathbb{C}^{N_r\times \tau}$ over the $\tau$  blocks in a coherence period is given as
\begin{equation}\label{eq:MIMO}
 \mathbf{Y}^{(t)} =  \sum\nolimits_{k=1}^K \mathbf{H}^{(t)}_k \bX_k^{(t)} + \bZ^{(t)}, 
\end{equation}
where $\bH_k^{(t)} \in\mathbb{C}^{N_r\times N_t}$ is the MIMO channel between the $k$-th device and the server, $\bX_k^{(t)} \in \mathbb{C}^{N_t\times \tau}$ is the transmit signal of device $k$, and $\bZ^{(t)}$  is the channel noise with identical and independent distributed elements following distribution $\cC\cN(0, N_0) $. As in \cite{zhong2023over}, we assume that the CSI $\{\bH_k^{(t)}\}_{k=1}^K$ is known to the server, and not each device $k$ has access to the local CSI $\bH_k^{(t)}$. The average transmit power constraints at each device $k$ across all antennas equals $P_0$.

\section{Low-Rank Gradient Factorization in Federated Learning}
\label{sec: algorithm}

In this section, we present the proposed over-the-air low-rank compression (Ota-LC) method.  Ota-LC reduces the communication overhead by imposing a low-rank factorization to subtensors of tensor $\bG^{(t)}$, which is known to serve as a useful inductive bias in many learning problems \cite{vogels2019powersgd}.  The approach applies to a wide range of neural network architectures, including multilayer perceptrons, convolutional networks, graph neural networks, and transformers. In fact, the high-dimension tensor of these model parameters can be partitioned into sub-tensors that can be reshaped into matrices suitable for low-rank factorization. For example, the gradient of a CNN takes the form of a 4-D tensor $\in \mathbb{R}^{O\times I \times H \times W}$, where $O$ represents the number of output channels, $I$ is the number of input channels, and $H\times W$ denotes the kernel size. For CNNs, one way to obtain sub-tensors is by partitioning along the dimension of the output channels \cite{vogels2019powersgd}.  

We start by proposing the design of distributed gradient factorization, followed by the implementation in MIMO systems \cite{marzetta_larsson_yang_ngo_2016} with error feedback aiming at a reduction of the bias of the low-rank compressed gradient  \cite{karimireddy2019error}.

\subsection{Ota-LC: Distributed LC for Gradient Factorization}\label{sec:compression}

We first formulate the gradient compression as a low-rank matrix factorization problem in a centralized system and then derive its distributed counterpart for FL.  With some abuse of notation, we write $\bG^{(t)}$ for any one of the subtensors in which the gradient is partitioned at round $t$. We reshape such subtensor into an $m\times n$ matrix, also denoted as $\bG^{(t)}\in\mathbb{R}^{m\times n}$. The high-dimensional global gradient matrix $\bG^{(t)}$ is represented by using low-rank matrices $\bP^{(t)}\in \mathbb{R}^{m\times r}$ and $\bQ^{(t)}\in \mathbb{R}^{n\times r}$  with rank  $r \ll \min(n,m)$, as $\bG^{(t)} = \bP^{(t)}[\bQ^{(t)}]^\top$. Finding the optimal matrix factorization is formulated as the minimization \cite{hastie2015matrix}
\begin{equation}\label{eq:low-rank}
    \min_{\bP^{(t)},\bQ^{(t)}} \frac{1}{2} \l\| \bP^{(t)}[\bQ^{(t)}]^\top - \bG^{(t)}\r\|_{F}^2 + \lambda(\| \bP^{(t)} \|_{F}^2 + \| \bQ^{(t)} \|_{F}^2),
\end{equation}
where $\lambda > 0$ is a regularization parameter, which controls the norms of factors $\bP^{(t)}$ and $\bQ^{(t)}$, and $\|\cdot \|_F$ denotes the Frobenius norm of a matrix. 
 The problem \eqref{eq:low-rank} is biconvex \cite{hastie2015matrix}, and a centralized algorithmic solution is given by multi-iteration Jacobi SCA \cite[Algorithm 1]{6675875}.

To elaborate, we introduce $s$ denoting Jacobi SCA iteration. Given initialization  $\bP^{(t,0)}$ and $\bQ^{(t,0)}$, for each iteration $s = 1,2,\ldots,S$, 
Jacobi SCA solves two surrogate convex problems
\begin{equation}
    \label{eq:surrogate function1}
  \min_{\bP} \frac{1}{2} \l\| \bP[\bQ^{(t,s-1)}]^\top - \bG^{(t)}\r\|_{F}^2 + \lambda\| \bP \|_{F}^2 ,
\end{equation}
and
\begin{equation}
    \label{eq:surrogate function2}
 \min_{\bQ} \frac{1}{2} \l\| \bP^{(t,s-1)}\bQ^\top - \bG^{(t)}\r\|_{F}^2 + \lambda \| \bQ \|_{F}^2
\end{equation}
in parallel. The closed-form optimal solution for problems \eqref{eq:surrogate function1} and \eqref{eq:surrogate function2} are 
\begin{equation}
\label{eq:als} 
    \begin{aligned}
        \overline{\bP}^{(t,s)} &= \bG^{(t)} \bQ^{(t,s-1)} {\l[({\bQ^{(t, s-1)}})^{\top} \bQ^{(t,s-1)} +\lambda \bI\r   ]}^{-1},\\
        \overline{\bQ}^{(t,s)} &= [\bG^{(t)}]^{\top} \bP^{(t,s-1)} {\l[({\bP^{(t,s-1)}})^{\top} \bP^{(t,s-1)} +\lambda \bI \r]}^{-1},
    \end{aligned}
\end{equation}
 for updating  $\bP^{(t,s)}$ and $\bQ^{(t,s)}$ as 
\begin{equation}
\label{eq:LLC update}
    \begin{aligned}
        \bP^{(t,s)} & = \bP^{(t,s-1)} + \beta^{(t,s)}(\overline{\bP}^{(t,s)} - \bP^{(t,s-1)}),\\
        \bQ^{(t,s)} & = \bQ^{(t,s-1)} + \beta^{(t,s)}(\overline{\bQ}^{(t,s)} - \bQ^{(t,s-1)}),
    \end{aligned}
\end{equation}
where  $\bI$ in \eqref{eq:als} is the identity matrix, and $\beta^{(t,s)} \in (0,1]$ is the step size of the update \eqref{eq:LLC update}.
At the last round $S$, we obtain an approximation of global gradient as $\bG^{(t)} \approx \bP^{(t,S)} [\bQ^{(t,S)}]^{\top}$. As a remark, when $\beta^{(t,s)}=1$, the update \eqref{eq:LLC update} is equivalent to the classical Jacobi best-response scheme, and its Gauss-Seidel counterpart update corresponds to alternating least squares (ALS)\cite{hastie2015matrix}. 

Consider now a federated learning setting. In this case,  by \eqref{eq: global gradient}, the gradient matrix $\bG^{(t)}$ is obtained as the sum $\bG^{(t)}=\sum_{k=1}^K\bG_k^{(t)}$, where matrix $\bG_k^{(t)}$ is only available at the $k$-th device. Plugging $\bG^{(t)} = \sum_{k=1}^K\bG_k^{(t)}$ into \eqref{eq:als}, we can write $\overline{\bP}^{(t,s)}$  and $     \overline{\bQ}^{(t,s)}$ in the update \eqref{eq:LLC update} as the sums
\begin{equation}\label{eq:aggregation}
         \overline{\bP}^{(t,s)} = \sum\nolimits_{k=1}^K   \overline{\bP}_k^{(t,s)} , \quad  \overline{\bQ}^{(t,s)} = \sum\nolimits_{k=1}^K \overline{\bQ}_k^{(t,s)}, 
\end{equation}
where the pair $\overline{\bP}_k^{(t,s)}\in\mathbb{R}^{m\times r}$ and $\overline{\bQ}_k^{(t,s)}\in\mathbb{R}^{n\times r}$ is evaluated at device $k$ with $\bG_k^{(t)}$ in lieu of $\bG^{(t)}$ via \eqref{eq:als}.

 This approach, however, requires a large communication and computation overhead, as low-rank factors  $\bP_k^{(t,s)}$ and ${\bQ}_k^{(t,s)}$ must be updated and exchanged across the rounds $s =1,2,\ldots,S$ of the LC update in \eqref{eq:LLC update} for each FL round $t$. Motivated by the observation that gradients typically exhibit strong temporal correlations across iterations \cite{azam2021recycling}, we address this issue by producing updated low-rank factors $\bP^{(t)}$ and $\bQ^{(t)}$ in a single update step \eqref{eq:LLC update}, i.e., $S=1$, and utilizing the approximation obtained at the last FL round $\bP^{(t-1)}$ and $\bQ^{(t-1)}$ as ``warm start" of the initial reference point. More precisely, the distributed version of the calculation \eqref{eq:als} is designed as
\begin{equation}
    \label{eq:ota_als_update}
    \begin{aligned}
        \overline{\bP}_k^{(t)} &= \bG_k^{(t)} \bQ^{(t-1)} {\l[({\bQ^{(t-1)}})^{\top} \bQ^{(t-1)} +\lambda \bI\r   ]}^{-1},\\
        \overline{\bQ}_k^{(t)} &= (\bG_k^{(t)})^{\top} \bP^{(t-1)} {\l[({\bP^{(t-1)}})^{\top} \bP^{(t-1)} +\lambda \bI \r]}^{-1},\\
    \end{aligned}
\end{equation}
for each round $t$, yielding the updated $\bP^{(t)}$ and $\bQ^{(t)}$ as
\begin{equation}
\label{eq:Ota-LC update}
    \begin{aligned}
        \bP^{(t)} & = \bP^{(t-1)} + \beta^{(t)}(\overline{\bP}^{(t)} - \bP^{(t-1)}),\\
        \bQ^{(t)} & = \bQ^{(t-1)} + \beta^{(t)}(\overline{\bQ}^{(t)} - \bQ^{(t-1)}),
    \end{aligned}
\end{equation}
where $\overline{\bP}^{(t)} = \sum_{k=1}^K \overline{\bP}_k^{(t)}$ and $\overline{\bQ}^{(t)} = \sum_{k=1}^K\overline{\bQ}_k^{(t)}$ as in \eqref{eq:aggregation}.

In Ota-LC, the local devices only need to upload two low-dimensional matrices $\overline{\bP}_k^{(t)}$ and $\overline{\bQ}_k^{(t)}$ instead of the original high-dimensional local gradient $\bG_k^{(t)}$. The communication overhead is thus reduced from $\mathcal{O}(n m)$ to $\mathcal{O}(nr+mr)$. In terms of computational complexity, the local updates \eqref{eq:ota_als_update} require an order $\mathcal{O}(nmr)$ of basic operation.

\subsection{Ota-LC: Implementation in MIMO Wireless System}\label{sec: signal design}

At each iteration, the devices transmit the updated local factorization matrices \eqref{eq:ota_als_update}  through the wireless MIMO channel using analog modulation. The server recovers the sum of the local factorizations from the received signal via over-the-air computation to obtain the low-rank factorizations of the global gradient. Ota-LC is summarized in Algorithm \ref{alg:OtA-LLC}, and a detailed implementation is provided next. 

\begin{algorithm}[htbp!]

 	\caption{Ota-LC with error feedback}
 	\label{alg:OtA-LLC}
 	\begin{algorithmic}[1]
            \State \textbf{Input}: Initialization $\bW^{(0)}$, $\bP^{(0)}$, $\bQ^{(0)}$, $\{\mathbf{G}_k^{(0)}\}_{k=1}^K$ and $\{\mathbf{\Delta}_k^{(0)}\}_{k=1}^K = 0$
 		\For {$t = 1,2,\ldots,T$}
           \State \parbox[t]{\dimexpr0.95\linewidth-\algorithmicindent}{Server calculates the receive beamforming matrix $\bA^{(t)}$ and the transmit beamforming matrices $\{\bB_k^{(t)}\}_{k=1}^K$ via \eqref{eq:beamformer}}
           \begin{spacing} {0.5}
               \State \parbox[t]{\dimexpr0.95\linewidth-\algorithmicindent}{Server broadcasts the low-rank factors $\bP^{(t-1)}$, $\bQ^{(t-1)}$ and $\bB_k^{(t)}$ to each device $k$}
           \end{spacing}
           
            \For{ each device $k = 1,2,\ldots,K$ (in parallel)} 
                \State Update the compression error $\mathbf{\Delta}_k^{(t-1)}$ via \eqref{eq:update error}
                \State\parbox[t]{\dimexpr0.9\linewidth-\algorithmicindent}{  Obtain the global gradient $\widehat{\bG}^{(t-1)}=\bP^{(t-1)}[\bQ^{(t-1)}]^\top$} 
                \State Update the the global model $\bW^{(t)}$ via \eqref{eq: Model update}
                \State\parbox[t]{\dimexpr0.9\linewidth-\algorithmicindent}{ Obtain gradient $\bG_k^{(t)}$ using gradient descent \eqref{eq: local gradient}}
                \State \parbox[t]{\dimexpr0.9\linewidth-\algorithmicindent}{Apply $\mathbf{\Delta}_k^{(t-1)}$ to obtain $\widetilde{\bG}_k^{(t)}$ in \eqref{eq:error feedback}}
                \State \parbox[t]{\dimexpr0.9\linewidth-\algorithmicindent}{Update the factors $\overline{\bP}_k^{(t)}$ and $\overline{\bQ}_k^{(t)}$  in \eqref{eq:ota_als_update}}
                \State Obtain the transmitted signal $\bS_k^{(t)}$ via \eqref{eq:matrix S}
                \State\parbox[t]{\dimexpr0.9\linewidth-\algorithmicindent}{ Precode the transmitted signal via \eqref{eq:transmitted signal} and transmit $\bX_k^{(t)}$ to server}
            \EndFor
            \State Server obtains the received signal $\bY^{(t)}$ as \eqref{eq:MIMO}
            \State\parbox[t]{\dimexpr0.95\linewidth-\algorithmicindent} {Server recovers the sum of transmitted signal $\widehat{\bS}^{(t)}$ via over-the-air computation beamforming as in \eqref{eq:beamforming}, and obtains $\widehat{\bP}^{(t)}$, $\widehat{\bQ}^{(t)}$}
            \begin{spacing}{0.3}
                \State\parbox[t]{\dimexpr0.95\linewidth-\algorithmicindent} {Server updates  $\bP^{(t)}$ and $\bQ^{(t)}$ via \eqref{eq:Ota-LC update}}
            \end{spacing}
        \EndFor
 	\end{algorithmic}
 \end{algorithm}

\subsubsection{Signal Processing}
In Ota-LC, the aggregation \eqref{eq:aggregation} is implemented using over-the-air computation to enhance spectrum efficiency. To start, we stack the obtained $\overline{\bP}_k^{(t)}$ and $\overline{\bQ}_k^{(t)}$ in \eqref{eq:ota_als_update} as the matrix 
\begin{equation}
\label{eq:matrix R}
    \bR_k^{(t)} = [(\overline{\bP}_k^{(t)})^\top,(\overline{\bQ}_k^{(t)})^\top ]^\top \in \mathbb{R}^{(m+n) \times r}.
\end{equation} 
At each device $k$, the matrix $\bR_k^{(t)}$ in \eqref{eq:matrix R}  is then arranged into a complex matrix {$\bC_{k}^{(t)}\in \mathbb{C}^{\frac{m+n}{2} \times r}$ } as 
\begin{equation}
    \label{eq:matrix C}
    \bC_{k}^{(t)}[i,:]=\bR_k^{(t)}[i,:] + j\bR_k^{(t)}[\frac{m+n}{2}+i,:], i = 1,\cdots ,\frac{m+n}{2}.
\end{equation}

The rank $r$ dictates the length of the communication blocks. In particular, with $N_t$ transmit antennas, the number of required channel uses in the time-frequency domain is 
\begin{equation}\label{eq: reduced rank}
     N_{cu} = {(m+n)r}/{(2N_t)} ,
\end{equation}
where all antennas transmit different entries of each $\bC_{k}^{(t)}$.  Then the matrix ${\bC}_k^{(t)}$ in \eqref{eq:matrix C}  is reshaped into $\bS_k^{(t)}$ as
 \begin{equation}
     \label{eq:matrix S}
     \bS_k^{(t)} = [\bc_{k,1}^{(t)},\ldots,\bc_{k,N_{cu}}^{(t)}]\in \mathbb{C}^{N_t\times N_{cu}},
 \end{equation}
 where $\bc_{k,b}^{(t)}=\bc_{k}^{(t)}((b-1)N_t + 1 : bN_t)\in \mathbb{C}^{N_t}, b = 1,\ldots,N_{cu}$, and $\bc_{k}^{(t)}\in \mathbb{C}^{N_t N_{cu}}$ is the obtained by vectorizing $\bC_{k}^{(t)}$.

Let $\bA^{(t)} \in \mathbb{C}^{N_r \times N_t}$ denote the receive beamforming matrix and let $\bB_k^{(t)} \in \mathbb{C}^{N_t\times N_t}$ denote the transmit beamforming matrix of device $k$ at iteration $t$. We assume that the server can accurately estimate the CSI of each device to obtain the beamforming matrices. We use the over-the-air beamforming method in \cite{zhu2018mimo}. The devices apply zero-forcing beamforming to transmit the local factorization matrices, while the server applies linear mean-squared error beamforming, yielding the beamforming matrices 
\begin{equation}\label{eq:beamformer}
    \begin{aligned}
    \bA^{(t)} &= \sqrt{\max_k \frac{1}{P_0}{\rm tr}\left( ({\bF^{(t)}})^\dag \bH_k^{(t)}[\bH_k^{(t)}]^\dag \bF^{(t)})^{-1} \right)}\bF^{(t)},\\
        \bB_k^{(t)} &= \left( (\bA^{(t)})^\dag\bH_k^{(t)} \right)^\dag \left( (\bA^{(t)})^\dag\bH_k^{(t)}[\bH_k^{(t)}]^\dag\bA^{(t)} \right)^{-1},
    \end{aligned}
\end{equation}
where $\bF^{(t)} = \sum_{k=1}^K\lambda_{\min}(\Sigma_k^2)\bU_k\bU_k^\dag \in\mathbb{C}^{N_r \times N_t}$, where $\bU_k$ and $\Sigma_k$ are obtained by compact singular value decomposition (SVD) of $\bH_k$ as $\bH_k=\bU_k\Sigma_k\bV_k^\dag$.
Then, the transmit signal of each device at each iteration $t$ is
\begin{equation}\label{eq:transmitted signal}
     \bX_k^{(t)} = \bB_k^{(t)} {\bS}_k^{(t)} \in \mathbb{C}^{N_t\times N_{cu}}.
 \end{equation}
 We assume the inequality, $N_{cu}\leq \tau$, which can be satisfied by choosing suitably the rank $r$.

At the server, by using the receive beamforming in \eqref{eq:MIMO}, the recovered signal is an approximation of the sum $\sum_{k=1}^K \bS_k^{(t)}$ given as  
 \begin{equation}
 \label{eq:beamforming}
			\widehat{\bS}^{(t)} = [{\bA^{(t)}}]^\dag \bY^{(t)} \in \mathbb{C}^{N_t\times \tau}.
\end{equation} 
Then, the server first reshapes matrix $\widehat{\bS}^{(t)}$ into $\widehat{\bC}^{(t)}\in\mathbb{C}^{\frac{m+n}{2}\times r}$ as an estimate of the sum $\sum_{k=1}^K \bC_k^{(t)}$, and it transforms matrix $\widehat{\bC}^{(t)}$ into real matrix $\widehat{\bR}^{(t)}\in\mathbb{R}^{(m+n)\times r}$, which is an estimate of the sum $\sum_{k=1}^K \bR_k^{(t)}$. The first $m$ columns of $\widehat{\bR}^{(t)}$ yield $\widehat{\bP}^{(t)}$, and the remaining columns yield  $\widehat{\bQ}^{(t)}$. The matrix $\widehat{\bP}^{(t)}$ and $\widehat{\bQ}^{(t)}$ are used for update in \eqref{eq:Ota-LC update} in lieu of $\overline{\bP}^{(t)}$ and $\overline{\bQ}^{(t)}$.

\subsubsection{Error feedback}

To alleviate the impact of the low-rank approximation error, Ota-LC introduces an error feedback mechanism in a manner similar to \cite{karimireddy2019error}. Accordingly, the local gradient $\bG_k^{(t)}$ is modified as 
\begin{equation}\label{eq:error feedback}
     \widetilde{\bG}_k^{(t)} = \bG_k^{(t)} + \mathbf{\Delta}_k^{(t-1)},
\end{equation}
where $\mathbf{\Delta}_k^{(t-1)}$ is the approximation error, which initialized as $\mathbf{\Delta}_k^{(0)} = 0$. 
The local gradient factorization step \eqref{eq:ota_als_update} is applied to the error-compensated gradient $ \widetilde{\bG}_k^{(t)} $. 
The compression error attached with device $k$ is set as
\begin{equation}
\label{eq:update error}
    \mathbf{\Delta}_k^{(t)} =  \widetilde{\bG}_k^{(t)} - \frac{1}{K}\bP^{(t)}[\bQ^{(t)}]^\top.
\end{equation}
By plugging \eqref{eq:error feedback} into \eqref{eq:update error}, one can see that error feedback attempts to compensate for compression errors, communication distortion, and client drifts, which are accumulated over the iterations.

 \section{Numerical Experiments}
 \label{sec: experiments}

We consider a wireless federated learning system with one server and $K=10$ edge devices.  In each communication round, only a randomly selected half of the devices upload their local gradients to the server. The transmit SNR $\rho = {P_0}/{N_0}$ is set to $20$ dB. The number of transmitter antennas of each device is set as $N_t=8$, and the number of the receive antennas of the server as $N_r=8$. The entries of the communication channels $\bH_k^{(t)}$ are  generated from i.i.d. $\mathcal{CN}(0,1)$ for all $k$ and $t$. The channels are assumed to be constant within each iteration $t$. The experiments were run with PyTorch on NVIDIA 3090 GPUs and the code can be found at \href{https://github.com/MingzhaoGuo/Ota-LC}{https://github.com/MingzhaoGuo/Ota-LC}.

The experiments are performed on the classification task for the Cifar-10 and Cifar-100 datasets in the i.i.d. and non-i.i.d. cases respectively. In the i.i.d. experiment, the data are randomly assigned to each user. In the non-i.i.d. experiment, the local datasets are divided using the Dirichlet distribution with scaling parameter $0.9$ \cite{hsu2019measuring}. 
We use the ResNet18 model, setting the learning rate as $\eta = 0.1$ for the Cifar-10 dataset and $\eta = 0.005$ for the Cifar-100 dataset, and the weight $\beta^{(t)}$ in LC update to $0.5$ for all $t$. We control the number of parameters uploaded by local devices by changing the value of the rank $r$.
Our method is compared with SGD without compression, as well as with five existing gradient compression methods with error feedback for wireless federated learning, as detailed next.
    
$\bullet$ Ota-CS \cite{zhong2023over}: Ota-CS employs Turbo-CS, a compressed sensing technique detailed in \cite{6883198}, to compress the transmitted gradients. Specifically, the method sparsifies the vectorized $mn\times 1$ gradients after projection into a subspace of dimension $(m+n)r$ via multiplication by an $(m+n)r \times mn$ partial discrete Fourier transform (DFT) compression matrix. This operation has a complexity of order $\mathcal{O}(nmr(m+n))$. This amounts to an increase of order $\mathcal{O}(m+n)$ as compared to the complexity of $\mathcal{O}(nmr)$ for Ota-LC. The server then reconstructs the compressed gradients using Turbo-CS. This method leverages over-the-air computation for data transmission.

$\bullet$ Ota-RLC\cite{xing2021federated}: In Ota-RLC, gradient compression employs RLC as introduced in \cite{abdi2020analog}. The local device first multiplies the vectorized gradients with a partial Hadamard matrix $\mathbf{H}_n\in \mathbb{R}^{(m+n)r \times mn} $, featuring mutually orthogonal rows, entailing a local computation cost of order $\mathcal{O}(nmr(m+n))$ as for Ota-CS. The server recovers the compressed gradient, using over-the-air computation for gradient transmission.

$\bullet$ PowerSGD\cite{vogels2019powersgd}: PowerSGD also compresses the local gradient by finding a low-rank approximation. Unlike Ota-LC, at each iteration, PowerSGD utilizes one step of power iteration, requiring two rounds of transmission and aggregation. The local computation has a complexity of order $\mathcal{O}(nmr)$, and PowerSGD leverages digital communications.  

 $\bullet$ Top-$K$\cite{alistarh2018convergence} and Rand-$K$\cite{stich2018sparsified}: The local computational cost of Top-$K$ has an order of $\mathcal{O}(mn\log(mn))$ due to the sorting, and Randk has an order of  $\mathcal{O}((m+n)r)$. These methods leverage digital communications.  

\begin{figure*}[!htbp]
    \centering
    \subfloat[]{
    \includegraphics[width = 0.9\columnwidth,trim = 0cm 0cm 0cm 0cm, clip]{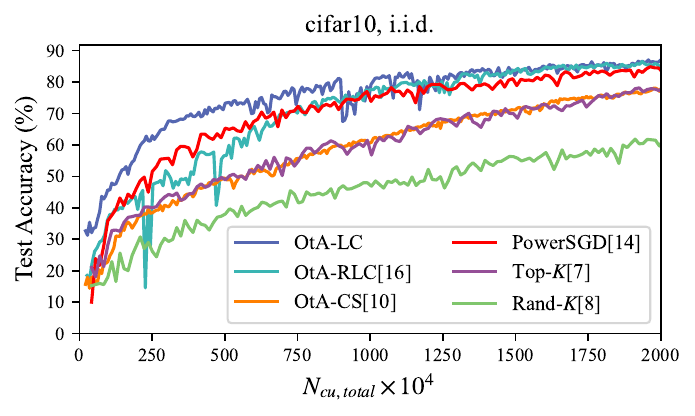}
    \label{fig:learning_curve(a)}
    }
    \subfloat[]{
    \includegraphics[width = 0.9\columnwidth,trim = 0cm 0cm 0cm 0cm, clip]{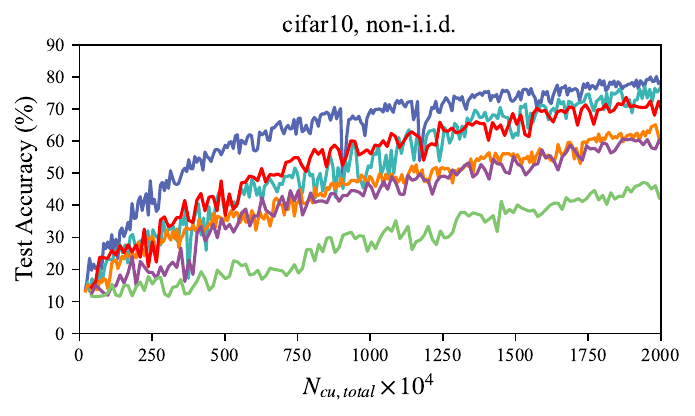}
    \label{fig:learning_curve(b)}
    }
    
    \subfloat[]{
    \includegraphics[width = 0.9\columnwidth,trim = 0cm 0cm 0cm 0cm, clip]{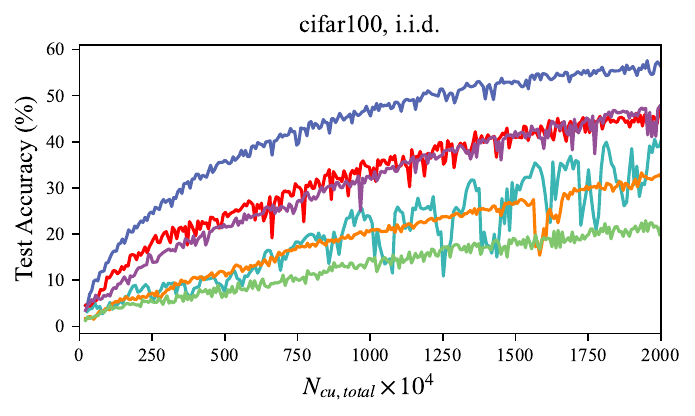}
    \label{fig:learning_curve(c)}
    }
    \subfloat[]{
    \includegraphics[width = 0.9\columnwidth,trim = 0cm 0cm 0cm 0cm, clip]{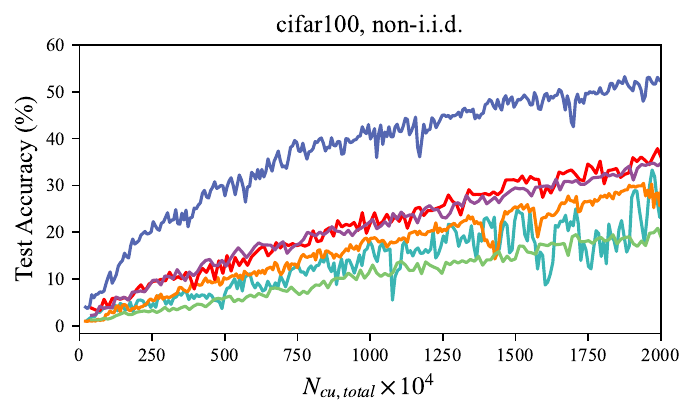}
    \label{fig:learning_curve(d)}
    }
    \setlength{\abovecaptionskip}{0.cm}
    \caption{Test accuracy as a function of the number of  $N_{cu,total}$ ($r = 20$).}
    \label{fig:learning_curve}
\end{figure*}

We first provide a convergence performance comparison by evaluating training loss and test accuracy against the total channel uses $N_{cu,total} = TN_{cu}$ and $r=20$.  It can be observed in Fig. \ref{fig:learning_curve} that Ota-LC demonstrates a faster convergence rate than other methods in both i.i.d. and non-i.i.d. scenarios. For instance, when aiming a learning accuracy of $70\%$ in Fig. \ref{fig:learning_curve(b)}, $1.5\times 10^7$ channel uses are required for Ota-RLC and PowerSGD, while Ota-LC requires only $1\times 10^7$ channel uses, reducing the communication overhead by $33\%$. 

\begin{figure}[htbp]
    \centering
    \subfloat[]{
    \includegraphics[width = 0.9\columnwidth]{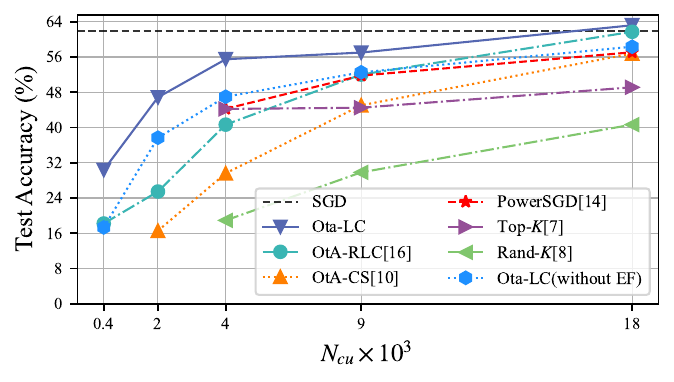}
    \label{fig:different compression(a)}
    }
    
    \subfloat[]{
    \includegraphics[width = 0.9\columnwidth]{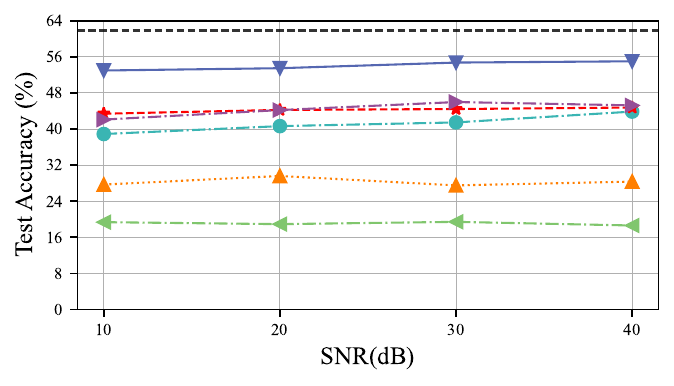}
    \label{fig:different compression(b)}
    }
    \setlength{\abovecaptionskip}{0.cm}
    \caption{ Test accuracy versus the number of $N_{cu}$ (a), and SNR (b)  ($T = 250$).} 
    \label{fig:different compression}
\end{figure}

 Then, we compare Ota-LC with the mentioned benchmarks for different numbers of channel uses $N_{cu}$ over $T=250$  and SNR $\rho = 20$ dB for Cifar-100 under the i.i.d. scenario.  The results in Fig. \ref{fig:different compression(a)} show that Ota-LC performs better under the same communication overhead, especially in the regime of a low number of communication resources $N_{cu}$,  and that error feedback helps improve performance. As $N_{cu}$ increases, the performance of Ota-LC gradually improves, approaching the performance of uncompressed gradient (SGD).
We also compare Ota-LC with the benchmarks for different SNR under $T=250$ and $N_{cu}=4000$.  Fig. \ref{fig:different compression(b)} shows that Ota-LC outperforms other algorithms even in low SNR regimes.

\section{Conclusion}
In this letter, we have proposed Ota-LC, a gradient compression and transmission method that reduces the communication overhead for MIMO wireless federated learning. Experimental results show that Ota-LC converges faster and has better performance than existing benchmarks while having smaller communication and computational costs. Future work may investigate the fully decentralized extension of Ota-LC and consider the effect of error feedback strength to enhance performance.

\bibliographystyle{IEEEtran}
\bibliography{reference}

\end{document}

%% file: header.tex
\def\phi{\varphi}

\def\l{\left}
\def\r{\right}
\def\({\left(}
\def\){\right)}

\def\bc{{\mathbf{c}}}

\def\bu{{\mathbf{u}}}

\def\b0{{\mathbf{0}}}

\def\bA{{\mathbf{A}}}
\def\bB{{\mathbf{B}}}
\def\bC{{\mathbf{C}}}

\def\bF{{\mathbf{F}}}
\def\bG{{\mathbf{G}}}
\def\bH{{\mathbf{H}}}
\def\bI{{\mathbf{I}}}

\def\bP{{\mathbf{P}}}
\def\bQ{{\mathbf{Q}}}
\def\bR{{\mathbf{R}}}
\def\bS{{\mathbf{S}}}

\def\bU{{\mathbf{U}}}
\def\bV{{\mathbf{V}}}
\def\bW{{\mathbf{W}}}
\def\bX{{\mathbf{X}}}
\def\bY{{\mathbf{Y}}}
\def\bZ{{\mathbf{Z}}}

\def\cC{\mathcal{C}}

\def\cN{\mathcal{N}}

